\documentclass{ws-ijgmmp}

\usepackage{graphicx}  
\usepackage{latexsym}
\usepackage{amssymb}
\usepackage{amsmath}
\usepackage{color}
\numberwithin{equation}{section}

\def\beq{\begin{equation}}
\def\eeq{\end{equation}}
\def\eqref#1{(\ref{#1})}

\begin{document}

\markboth{D. Bini and B. Mashhoon}
{Nonlocal gravity: conformally flat spacetimes}

%
\catchline{}{}{}{}{}
%

\title{NONLOCAL GRAVITY: CONFORMALLY FLAT SPACETIMES
}

\author{\footnotesize DONATO BINI}

\address{
Istituto per le Applicazioni del Calcolo ``M. Picone,'' CNR, I--00185 Rome, Italy\\
INFN - Sezione di Napoli, Complesso Universitario di Monte S. Angelo, Via Cintia Edificio 6, 80126 Napoli, Italy\\
\email{donato.bini@gmail.com}
}

\author{\footnotesize BAHRAM MASHHOON}

\address{
Department of Physics and Astronomy,\\
University of Missouri, Columbia, Missouri 65211, USA\\
\email{MashhoonB@missouri.edu}
}

\maketitle

\begin{history}
\received{(Day Month Year)}
\revised{(Day Month Year)}
\end{history}

\begin{abstract}
The field equations of the recent nonlocal generalization of Einstein's theory of gravitation are presented in a form that is reminiscent of general relativity. The implications of the nonlocal  field equations are studied in the case of conformally flat spacetimes. Even in this simple case, the field equations are intractable. Therefore, to gain insight into the nature of these equations, we investigate the structure of nonlocal gravity in two-dimensional spacetimes. While any smooth 2D spacetime is conformally flat and satisfies Einstein's field equations,  
only a subset containing either a Killing vector or a homothetic Killing vector can satisfy the field equations of nonlocal gravity. 
\end{abstract}

\begin{flushleft}
\end{flushleft}

\section{Introduction}

In Minkowski spacetime, Lorentz invariance is extended to accelerated observers on the basis of the assumption that an accelerated observer is pointwise inertial~\cite{Ei}. That is, at each instant along its world line, an accelerated observer is supposed to be momentarily equivalent to an otherwise identical hypothetical comoving inertial observer. According to this \emph{hypothesis of locality}, the accelerated observer thus passes through a continuous infinity of momentarily comoving inertial observers. To determine the measurements of the accelerated observer, pointwise Lorentz transformations can be employed in conformity with the locality postulate~\cite{Mash1, Mash2, Mash3}. The locality postulate plays a crucial role in general relativity (GR) as well. To extend relativity theory to the gravitational domain, Einstein postulated a certain local connection between an observer in a gravitational field and a corresponding accelerated observer in Minkowski spacetime. The latter is locally inertial by the hypothesis of locality. Therefore, Einstein's principle of equivalence together with the hypothesis of locality renders observers in a gravitational field locally inertial~\cite{Ei}. 

The special theory of relativity can naturally treat pointlike coincidences of classical particles and rays of radiation, since such interactions are consistent with the locality postulate. On the other hand, classical electromagnetic field measurements are intrinsically nonlocal in accordance with the Bohr-Rosenfeld principle~\cite{BoRo, BoRo2}. In 1933, Bohr and Rosenfeld pointed out that only spacetime averages of classical electric and magnetic fields have physical significance, since these fields cannot be measured instantaneously~\cite{BoRo}. To go beyond the locality postulate for field measurements, one must therefore take due account of the past \emph{history} of an accelerated observer.  On this basis, a \emph{nonlocal special relativity theory} has been developed, where partial differential equations of the electromagnetic field, for instance, have been replaced by partial integro-differential equations, see Ref.~\cite{Mash4} and the references cited therein.  

The fundamental connection between \emph{inertia} and \emph{gravitation}, originally elucidated by Einstein~\cite{Ei}, provides the motivation to seek a history-dependent generalization of Einstein's theory of gravitation. In such a theory, the gravitational field would be local, but would satisfy partial integro-differential  field equations that would reduce to Einstein's field equations in the local limit. Einstein's general relativity (GR) is a field theory modeled after electrodynamics. Maxwell's original equations involve field tensors $F_{\mu\nu} \mapsto (\mathbf{E}, \mathbf{B})$ and $H_{\mu\nu} \mapsto (\mathbf{D}, \mathbf{H})$ and the constitutive relation between these field quantities is in general nonlocal~\cite{Jack}.  It turns out that GR has an equivalent formulation, GR$_{||}$, within the framework of teleparallelism. Indeed, GR$_{||}$, the teleparallel equivalent of GR,  is the gauge theory of the Abelian group of spacetime translations. Therefore, the field equations of GR$_{||}$ bear a formal resemblance to Maxwell's original equations of electrodynamics of media. Thus GR$_{||}$ can be rendered nonlocal in analogy with electrodynamics of media via the introduction of a scalar causal constitutive kernel~\cite{HeMa, HeMa2}. In this way, one arrives at an indirect nonlocal generalization of GR. 

At the present stage of the development of classical nonlocal gravity, the constitutive kernel of the theory must be determined from observational data. Moreover, the nonlocal aspect of the gravitational interaction in this theory appears to simulate \emph{dark matter}~\cite{NL8}. That is, there is no dark matter in nonlocal gravity; however, what appears as dark matter in astronomy may be due to the nonlocality of the gravitational interaction. 

The field equations of nonlocal gravity (NLG) are~\cite{NL9, NL10}
\begin{equation}\label{I1}
 {^0}G_{\mu \nu}+\Lambda\, g_{\mu \nu}+{\cal N}_{\mu \nu}-Q_{\mu \nu}=\kappa\, T_{\mu \nu}\,.
  \end{equation}
In our convention, an event in spacetime has coordinates $x^\mu=(ct, x^i)$, where Greek indices run from 0 to 3, while Latin indices run from 1 to 3; moreover, the spacetime metric has signature +2, $\kappa:=8 \pi G/c^4$ and $c=1$, unless otherwise specified. The gravitational field equations in general relativity  (GR) are given by 
\begin{equation}\label{I2}
  {^0}G_{\mu \nu}+  \Lambda\, g_{\mu \nu}= \kappa~T_{\mu \nu}\,, \qquad  {^0}G_{\mu \nu}={^0}R_{\mu \nu}-\frac{1}{2} g_{\mu \nu}\,{^0}R\,,
 \end{equation}
where ${^0}R_{\mu \nu}={^0}R^{\alpha}{}_{\mu \alpha \nu}$ is the symmetric Ricci tensor, ${^0}R= g^{\mu \nu}~{^0}R_{\mu \nu}$ is the scalar curvature, $\Lambda$ is the cosmological constant and $T_{\mu \nu}$ is the \emph{symmetric} energy-momentum tensor of matter. This is the standard differential formulation of GR; however, we mention in passing that it is possible to provide an integral formulation of GR as well~\cite{Sciama:1970yk}.
Comparing Eqs.~\eqref{I1} and~\eqref{I2}, we notice the existence of two extra terms of nonlocal origin, namely, ${\cal N}_{\mu \nu}$ and $Q_{\mu \nu}$, which are not in general symmetric tensors. In fact, ${\cal N}_{\mu \nu}$  is given by
\begin{equation}\label{I3}
{\cal N}_{\mu \nu}:=g_{\nu \alpha}\, e_\mu{}^{\hat{\gamma}}\,\frac{1}{\sqrt{-g}}\,\frac{\partial}{\partial x^\beta}\,\Big(\sqrt{-g}\,N^{\alpha \beta}{}_{\hat{\gamma}}\Big)\,,
\end{equation}
where $N^{\alpha \beta}{}_{\hat{\gamma}}=e_{\mu \hat{\gamma}}\,N^{\alpha \beta \mu}$ and $Q_{\mu \nu}$ is a traceless tensor (in 4D) that can be expressed as
\begin{equation}\label{I4}
Q_{\mu \nu}:=C_{\mu \rho \sigma} N_{\nu}{}^{\rho \sigma}-\frac 14\, g_{\mu \nu}\,\Sigma\,, \qquad \Sigma=\,C_{ \delta \rho \sigma}N^{\delta \rho \sigma}\,.
\end{equation}
It remains to specify the nonlocal aspects of the field equations of NLG and the  tensor $N_{\alpha \beta \gamma}(x)$. To this end, the usual GR framework has to be extended; therefore, a brief digression is necessary. 

We work in an extended GR framework, where the spacetime is occupied by a \emph{preferred} set of observers with orthonormal tetrad frames $e^\mu{}_{\hat{\alpha}}$, which carry the gravitational degrees of freedom. Indeed, in Eq.~\eqref{I1}, we have sixteen field equations for the sixteen components of the gravitational potentials specified by our preferred tetrad field $e_\mu{}^{\hat{\alpha}}$. The pseudo-Riemannian spacetime metric has Lorentzian signature and is given by
\begin{equation}\label{I5}
 g_{\mu \nu}(x)= \eta_{\hat {\alpha} \hat  {\beta}}\,
e_\mu{}^{\hat {\alpha}}(x)~e_\nu{}^{\hat {\beta}}(x)\,. 
\end{equation}
Starting from the sixteen components of the tetrad frame field of our preferred observers, the ten orthonormality conditions for the frame field are equivalent to the definition of the metric in terms of the frame field~\eqref{I5}.  
The preferred observers' tetrad frames lead to the linear \emph{Weitzenb\"ock connection}~\cite{We}
\begin{equation}\label{I6}
\Gamma^\mu_{\alpha \beta}=e^\mu{}_{\hat{\rho}}~\partial_\alpha\,e_\beta{}^{\hat{\rho}}\,,
\end{equation}
which corresponds to a covariant differentiation such that 
\begin{equation}\label{I7}
\nabla_\nu\,e_\mu{}^{\hat{\alpha}}=0\,.
\end{equation}

The nonsymmetric Weitzenb\"ock  connection is curvature free and renders the preferred observers'  frame field parallel; moreover, it is compatible with the spacetime metric, since  $\nabla_\nu\, g_{\alpha \beta}=0$ follows from Eqs.~\eqref{I5} and~\eqref{I7}. 

The extended GR framework thus has a pseudo-Riemannian metric and two metric-compatible linear connections: The Levi-Civita $(^{0}\Gamma^\mu_{\alpha \beta})$ connection and the  Weitzenb\"ock $(\Gamma^\mu_{\alpha \beta})$ connection. In our convention, a left superscript ``0" is used throughout to denote geometric quantities in GR that are directly related to the Levi-Civita connection.   The difference between two linear connections on the same spacetime manifold is a tensor; thus, we have the \emph{torsion} tensor 
\begin{equation}\label{I8}
 C_{\alpha \beta}{}^{\mu}=\Gamma^{\mu}_{\alpha \beta}-\Gamma^{\mu}_{\beta \alpha}=e^\mu{}_{\hat{\rho}}\Big(\partial_{\alpha}e_{\beta}{}^{\hat{\rho}}-\partial_{\beta}e_{\alpha}{}^{\hat{\rho}}\Big)\,
\end{equation}
and the \emph{contorsion} tensor
\begin{equation}\label{I9}
K_{\alpha \beta}{}^\mu= {^0} \Gamma^\mu_{\alpha \beta} - \Gamma^\mu_{\alpha \beta}\,.
\end{equation}
The torsion tensor is antisymmetric in its first two indices, while the contorsion tensor is antisymmetric in its last two indices. It follows directly from the metric compatibility of our two connections that
\begin{equation}\label{I10}
K_{\alpha \beta \gamma} = \frac{1}{2} (C_{\alpha \gamma \beta}+C_{\beta \gamma \alpha}-C_{\alpha \beta \gamma})\,.
\end{equation}
Furthermore, it proves useful to introduce an auxiliary torsion tensor
\begin{equation}\label{I11}
\mathfrak{C}_{\alpha \beta \gamma} :=K_{\gamma \alpha \beta}+C_\alpha\, g_{\gamma \beta} - C_\beta \,g_{\gamma \alpha}\,,
\end{equation}
where $C_\alpha:=- C_{\alpha}{}^{\beta}{}_{\beta}$ is the \emph{torsion covector}. Then
\begin{eqnarray}\label{I12}
N_{\mu \nu \rho} = - \int \Omega_{\mu \mu'} \Omega_{\nu \nu'} \Omega_{\rho \rho'}\, {\cal K}(x, x')\,X^{\mu' \nu' \rho'}(x') \sqrt{-g(x')}\, d^4x' \,,
\end{eqnarray}
where ${\cal K}(x, x')$ is the scalar \emph{causal} kernel of nonlocal gravity~\cite{HeMa, HeMa2,NL8, NL9, NL5} and  $X_{\mu \nu \rho}=-X_{\nu \mu \rho}$ depends upon the spacetime torsion. To ensure causality, we assume that  event $x'$ is connected  to event $x$ by means of a unique future directed timelike or null geodesic and we define the square of the proper length of this geodesic to be $2\,\Omega$, where $\Omega$ is the \emph{world function}~\cite{Hadam,Ruse:1931ht,Synge:1931zz,Sy}. The  indices $\mu', \nu', \rho',...$, in Eq.~\eqref{I12} refer to event $x'$, while indices $\mu, \nu, \rho, ...$, refer to event $x$. Furthermore, we define covectors
\begin{equation}\label{I13}
\Omega_{\mu}(x, x'):=\frac{\partial \Omega}{\partial x^{\mu}}, \quad \Omega_{\mu'}(x, x'):=\frac{\partial \Omega}{\partial x'^{\mu'}}\,,
\end{equation}
which are tangent to the geodesic at $x$ and $x'$, respectively, see Figure 1.  It can be shown that for any bitensor, the covariant derivatives at $x$ and $x'$ commute~\cite{Sy}. Indeed,  $\Omega_{\mu \mu'}(x, x')=\Omega_{\mu' \mu}(x, x')$ is a dimensionless bitensor that essentially reduces to the metric tensor in the local limit; that is, 
\begin{equation}\label{I14}
\lim_{x' \to x} \Omega_{\mu \mu'}(x, x')=-g_{\mu \mu'}(x)\,.
\end{equation}

It remains to specify how $X_{\mu \nu \rho}$ is related to the torsion tensor. In analogy with electrodynamics~\cite{HO}, we assume  a \emph{linear} relation, i.e., 
\begin{equation}\label{I15}
X_{\mu \nu \rho}=\chi_{\mu \nu \rho}{}^{\alpha \beta \gamma}\,\mathfrak{C}_{\alpha \beta \gamma}\,.
\end{equation}
A detailed examination of possible forms of Eq.~\eqref{I15} is contained in  Ref.~\cite{NL9}. In NLG, we adopt the linear relation 
\begin{equation}\label{I16}
X_{\mu \nu \rho}= \mathfrak{C}_{\mu \nu \rho}+ \hat{p}\,(\check{C}_\mu\, g_{\nu \rho}-\check{C}_\nu\, g_{\mu \rho})\,,
\end{equation}
where $\hat{p}\ne 0$ is a constant dimensionless parameter and $\check{C}^\mu$ is the torsion pseudovector given by
\begin{equation}\label{I17}
\check{C}^\alpha=\frac{1}{3} E^{\alpha}{}_{\beta \gamma \delta}\,\mathfrak{C}^{\beta \gamma \delta}\,.
\end{equation}
Here, $E_{\alpha \beta \gamma \delta}$ is the Levi-Civita tensor. 

The field equations of NLG have been studied thus far only in the linear approximation for which it is sufficient to employ the world function $\Omega$ in the limit of Minkowski spacetime, i.e., 
\begin{equation}\label{I18}
^M\Omega= \frac{1}{2} \eta_{\mu \nu}\, (x^\mu-x'^\mu)\,(x^\nu-x'^\nu)\,.
\end{equation}

We emphasize that Eq.~\eqref{I12} is the simplest relation of its kind, since we use a \emph{scalar} kernel; moreover, the kernel is \emph{causal}, since event $x$ is in the future of event $x'$. That is, Eq.~\eqref{I12} represents a certain average over the past state of the gravitational field. The causal kernel is expected to be a function of the relevant scalars in the problem such as
\begin{equation}\label{I18a}
\Omega_\mu\, e^\mu{}_{\hat{\alpha}}(x)\,, \qquad   \Omega_{\mu'}\, e^{\mu'}{}_{\hat{\alpha}}(x')\,.
\end{equation}
In linearized NLG, we find that ${\cal K}(x, x')$ is a convolution kernel of the form 
\begin{equation}\label{I18b}
{\cal K}(x, x')=\Theta(t-t'-|\mathbf{x}-\mathbf{x'}|) \, k(x-x')\,,
\end{equation}
where $\Theta$ is the unit step function such that $\Theta(t)=0$ for $t<0$ and $\Theta(t)=1$ for $t\ge0$.

Let us now return to Eq.~\eqref{I1} and define ${\cal T}_{\mu \nu}$ in terms of the nonlocal parts of the field equations of NLG, namely,
\begin{equation}\label{I19}
{\cal T}_{\mu \nu}= \kappa^{-1}\,(Q_{\mu \nu}-{\cal N}_{\mu \nu})\,,
\end{equation}
so that Eq.~\eqref{I1} can now be written as 
\begin{equation}\label{I20}
 {^0}G_{\mu \nu}+\Lambda\, g_{\mu \nu}=\kappa\, (T_{\mu \nu}+ {\cal T}_{\mu \nu})\,.
\end{equation}
Here ${\cal T}_{(\mu \nu)}$ has the interpretation of the symmetric energy-momentum tensor of the effective dark matter, while  
\begin{equation}\label{I21}
 {\cal T}_{[\mu \nu]}=0\,
\end{equation}
are the six constraint equations that are necessary in order to determine the sixteen components of the tetrad frame field of the preferred observers. That is, the field equations of NLG consist of the ten nonlocally modified Einstein equations
\begin{equation}\label{I22}
 {^0}G_{\mu \nu}+\Lambda\, g_{\mu \nu}=\kappa\, [T_{\mu \nu}+ {\cal T}_{(\mu \nu)}]\,.
\end{equation}
together with the six constraint equations~\eqref{I21}. Thus in this theory what appears in astrophysics and cosmology as dark matter is in fact the nonlocal aspect of the gravitational interaction. Furthermore, it follows from the reduced Bianchi identity, $^0\nabla_\nu\,^0G^{\mu \nu}=0$, that the \emph{total} matter energy-momentum tensor is conserved, namely, 
\begin{equation}\label{I23}
 {^0}\nabla_\nu \left[T^{\mu \nu}+{\cal T}^{(\mu \nu)}\right]=0\,.
\end{equation}

It is interesting to investigate gravitational systems that consist entirely of effective dark matter $(T_{\mu \nu}=0)$, which, based on previous work in the linear regime, would lack a proper Newtonian limit and hence could exist only in highly relativistic situations, as well as systems for which nonlocal effects vanish. Imagine the latter circumstance, where the preferred observers' tetrad frames in Eq.~\eqref{I1} are such that the corresponding metric tensor $g_{\mu \nu}$ satisfies Eq.~\eqref{I2}. Then, ${\cal T}_{\mu \nu}=0$ and ${\cal N}_{\mu \nu}=Q_{\mu \nu}$ in this case. It follows from the trace free nature of $Q_{\mu \nu}$ in 4D  that 
$g^{\mu \nu}\,{\cal N}_{\mu \nu}=0$, which means that 
\begin{equation}\label{I24}
 e_\alpha{}^{\hat{\gamma}}\,\frac{1}{\sqrt{-g}}\,\frac{\partial}{\partial x^\beta}\,\Big(\sqrt{-g}\,N^{\alpha \beta}{}_{\hat{\gamma}}\Big)=0\,.
\end{equation}

To find a solution of NLG, we must ultimately determine the sixteen components of the tetrad frame field $e^\mu{}_{\hat{\alpha}}$ of the preferred observers. Of these spacetime functions, ten would then specify the metric tensor in accordance with Eq.~\eqref{I5}. Let us note that with $e^\mu{}_{\hat{\alpha}}=\delta^\mu_{\hat \alpha}$, $g_{\mu \nu}=\eta_{\mu \nu}$, $\Lambda=0$ and $T_{\mu \nu}=0$, we find that the tetrads of the ideal inertial observers as the preferred observers in Minkowski spacetime constitute an exact solution of NLG. \emph{No other exact solution of NLG is known}, because of the complicated structure of nonlocal gravity theory. Thus far, only NLG theory that is \emph{linearized about ideal inertial observers in Minkowski spacetime} has been investigated~\cite{NL9}. In these studies, it is sufficient to employ the world function of Minkowski spacetime, which enormously simplifies the task of finding solutions of the theory. Otherwise, the world function for the timelike geodesics of curved spacetime is required in the nonlocal ansatz~\eqref{I12} and the causal scalar kernel ${\cal K}(x,x')$ should be determined from the comparison of the theory with observation. 

To go one step beyond Minkowski spacetime and explore the structure of NLG in the nonlinear regime, we consider preferred observers in conformally flat spacetimes.

\section{Conformally Flat Spacetimes}

Consider a metric of the form
\begin{equation}\label{2.1}
ds^2= e^{2U} \, \eta_{\mu \nu}\, dx^\mu\,dx^\nu\,,
\end{equation}
where $U(x)$ is a scalar under general coordinate transformations. We choose preferred observers that are at rest in space with 
\begin{equation}\label{2.1a}
e^\mu{}_{\hat{\alpha}}=e^{-U}\,\delta^\mu_{\hat{\alpha}}\,, \qquad     e_\mu{}^{\hat{\alpha}}=e^{U}\,\delta_\mu^{\hat{\alpha}}\,.
\end{equation}

It is then straightforward to show that the torsion tensor is
\begin{equation}\label{2.2}
 C_{\alpha \beta}{}^{\mu}=U_{\alpha}\, \delta^\mu_\beta- U_{\beta}\, \delta^\mu_\alpha\,, \quad  C_{\alpha \beta \gamma}=e^{2U}\,(U_{\alpha}\, \eta_{\beta \gamma}- U_{\beta}\, \eta_{\gamma \alpha})\,,
\end{equation}
where $U_{\mu}:= \partial_{\mu}U$ in this case, and the torsion covector is
\begin{equation}\label{2.3}
 C_{\alpha}=-3\,U_{\alpha}\,. 
\end{equation}
Moreover, the contorsion tensor is given by
\begin{equation}\label{2.4}
 K_{\alpha \beta \gamma}=C_{\beta\alpha\gamma}
\end{equation}
and the auxiliary torsion tensor is
\begin{equation}\label{2.5}
 \mathfrak{C}_{\alpha \beta \gamma}=-2\,C_{\alpha \beta \gamma}\,. 
\end{equation}
It follows that in general $\check{C}_\alpha=0$ in this case. Hence, $X_{\mu \nu \rho}= \mathfrak{C}_{\mu \nu \rho}$. 

For conformally flat spacetimes, the Einstein tensor is given by~\cite{Ste}
\begin{equation}\label{2.6}
^0G_{\mu \nu}=-2\,(U_{\mu \nu}-U_\mu\,U_\nu)+\eta_{\mu \nu}(U^\alpha\,U_\alpha+2\,U^\alpha{}_\alpha)\,,
\end{equation}
where $U^{\alpha} := \eta^{\alpha \beta}U_\beta$,  $U^\alpha{}_\beta :=\partial_{\beta}U^{\alpha}$, etc. Moreover, in Eq.~\eqref{I1}, the matter content could be due to a perfect fluid of energy density $\rho$ and pressure $p$ such that 
\begin{equation}\label{2.7}
T_{\mu \nu}=\rho\,u_\mu\,u_\nu+p\,(g_{\mu \nu}+u_\mu\,u_\nu)\,,
\end{equation}
where $u^\mu$ is the 4-velocity vector of the perfect fluid. The nonlocal parts of Eq.~\eqref{I1} can be expressed in this case as 
\begin{equation}\label{2.8}
{\cal N}_{\mu \nu}=e^{-U} \,\eta_{\nu \alpha}\,\frac{\partial}{\partial x^\beta}\,\Big(e^{3\,U}\,N^{\alpha \beta}{}_{\mu}\Big)\,,
\end{equation}
and 
\begin{equation}\label{2.9}
Q_{\mu \nu}=U_\mu\,N_{\nu}{}^{\rho}{}_\rho-U_\rho\,N_{\nu}{}^{\rho}{}_\mu-\frac{1}{4}\,g_{\mu \nu}\,(U_\alpha\,N^{\alpha \beta}{}_{\beta}-U_\beta\,N^{\alpha \beta}{}_{\alpha})\,,
\end{equation}
where
\beq\label{2.10}
N_{\mu \nu \rho}(x) = 2\, \int (\Omega_{\mu}{}^{\mu'}\,U_{\mu'}\, \Omega_{\nu \rho'}-\Omega_{\nu}{}^{\nu'}\,U_{\nu'}\, \Omega_{\mu \rho'})\,\Omega_{\rho}{}^{\rho'}\, {\cal K}(x, x')\,e^{4\,U(x')}\, d^4x' \,.
\eeq
It is clear that even in this simple case, the field equations of NLG are extremely complicated; moreover, to go forward, we require an explicit expression for the world function. 

\section{World Function}

We assume that two causally separated events are connected by a {\it
  unique} timelike or null geodesic; more generally, in the spacetime
region under consideration, there exists a unique geodesic joining
every pair of events. It then proves useful to employ the
{\it world function} $\Omega$, which denotes half the square of the proper
distance from $P':x'=\xi(\zeta_0)$ to $P:x=\xi(\zeta_1)$ along the
geodesic path $x^\alpha=\xi^\alpha(\zeta)$, see Figure 1. That is, we define \cite{Sy}
\begin{equation}\label{world}
  \Omega(x,x')=\frac 12 (\zeta_1-\zeta_0)\int_{\zeta_0}^{\zeta_1}
  g_{\mu \nu}\frac{d\xi^\mu}{d\zeta}\frac{d\xi^\nu}{d\zeta}d\zeta\,.
\end{equation}
It turns out that $\Omega$ is independent of the affine parameter
$\zeta$; moreover, the integrand in Eq.~(\ref{world}) is constant by
virtue of the geodesic equation. The main properties of $\Omega(x,x')$ are
summarized below.

Consider a variation of Eq.~(\ref{world}) that changes the endpoints, then
\begin{equation}\label{deltaOmega}
\delta\Omega(x,x')=(\zeta_1-\zeta_0)\left[g_{\mu \nu}\frac{d\xi^\nu}{d\zeta}
\delta\xi^\mu \right]_{\zeta_0}^{\zeta_1}\,.
\end{equation}
On the other hand,
\begin{equation}\label{d01}
\delta \Omega=\frac{\partial\Omega}{\partial x^\alpha}\delta x^\alpha+
\frac{\partial\Omega}{\partial x'^{\alpha'}}\delta x'^{\alpha'}\,,
\end{equation}
so that
\begin{eqnarray}\label{d02}
  \frac{\partial\Omega}{\partial
    x^\alpha}&=& \nonumber (\zeta_1-\zeta_0)g_{\alpha \beta}(x)\frac{dx^\beta}
{d\zeta}\,,\\\frac{\partial\Omega}{\partial x'^{\alpha'}}&=&-
(\zeta_1-\zeta_0)g_{\alpha' \beta'}(x')\frac{dx'^{\beta'}}{d\zeta}\,.
\end{eqnarray}
It is possible to see from the geodesic equation that the integrand in
Eq.~(\ref{world}) is indeed constant; therefore,
\begin{eqnarray}\label{Omega1}
  \Omega(x,x')&=& \nonumber \frac 12(\zeta_1-\zeta_0)^2 
g_{\mu \nu}(x)\frac{dx^\mu}{d\zeta}
  \frac{dx^\nu}{d\zeta}\\ &=&\frac 12(\zeta_1-\zeta_0)^2
  g_{\mu' \nu'}(x')\frac{dx'^{\mu'}}
  {d\zeta}  \frac{dx'^{\nu'}}{d\zeta}\,.
\end{eqnarray}
It follows from Eqs.~(\ref{d02})--(\ref{Omega1}) that
\begin{equation}\label{Omega2}
2\Omega=g^{\mu \nu}\Omega_\mu\Omega_\nu=g^{\mu' \nu'}\Omega_{\mu'}\Omega_{\nu'}\,,
\end{equation}
see Figure 1. Moreover, $\Omega=0$ for a null geodesic, $\Omega=-\frac
12\,\tau^2$ for a timelike geodesic of length $\tau$ and
$\Omega=\frac 12\,\sigma^2$ for a spacelike geodesic of length $\sigma$.

Let $k^\mu$ be a Killing vector field; then, $k_\mu\,dx^\mu/d\tau$ is a constant along a geodesic path. It follows from this fact and Figure 1 that 
\begin{equation}\label{Omega3}
k^\mu(x)\,\Omega_\mu(x, x')+ k^{\mu'}(x')\,\Omega_{\mu'}(x, x')=0\,.
\end{equation}

In Minkowski spacetime, $^M\Omega$ is given by Eq.~\eqref{I18}, so that
\begin{equation}\label{OmegaMink1}
^M\Omega(x,x')=-\frac 12\left[(t'-t)^2-(\mathbf{x'}-\mathbf{x})^2 \right]\,,
\end{equation}
since, in accordance with our convention, $\eta_{\alpha \beta}={\rm diag}\,(-1,1,1,1)$.
In this case, we find that
\begin{equation}\label{OmegaMink2}
^M\Omega_{\mu \mu'}=\frac{\partial^2\,^M\Omega(x,x')}{\partial x^\mu\partial x'^{\mu'}}=
-\eta_{\mu \mu'}\,,
\end{equation}
while
\begin{equation}\label{OmegaMink3}
 ^M\Omega_{\mu \nu}=\frac{\partial^2\,^M\Omega}{\partial x^\mu\partial x^\nu}=
  \eta_{\mu \nu}\,,\quad ^M\Omega_{\mu' \nu'}=\frac{\partial^2\,^M\Omega}{\partial x'^{\mu'}
\partial x'^{\nu'}}=\eta_{\mu' \nu'}\,.
\end{equation}

\begin{figure}
\begin{center}
\[
\begin{array}{cc}
\includegraphics[scale=0.35]{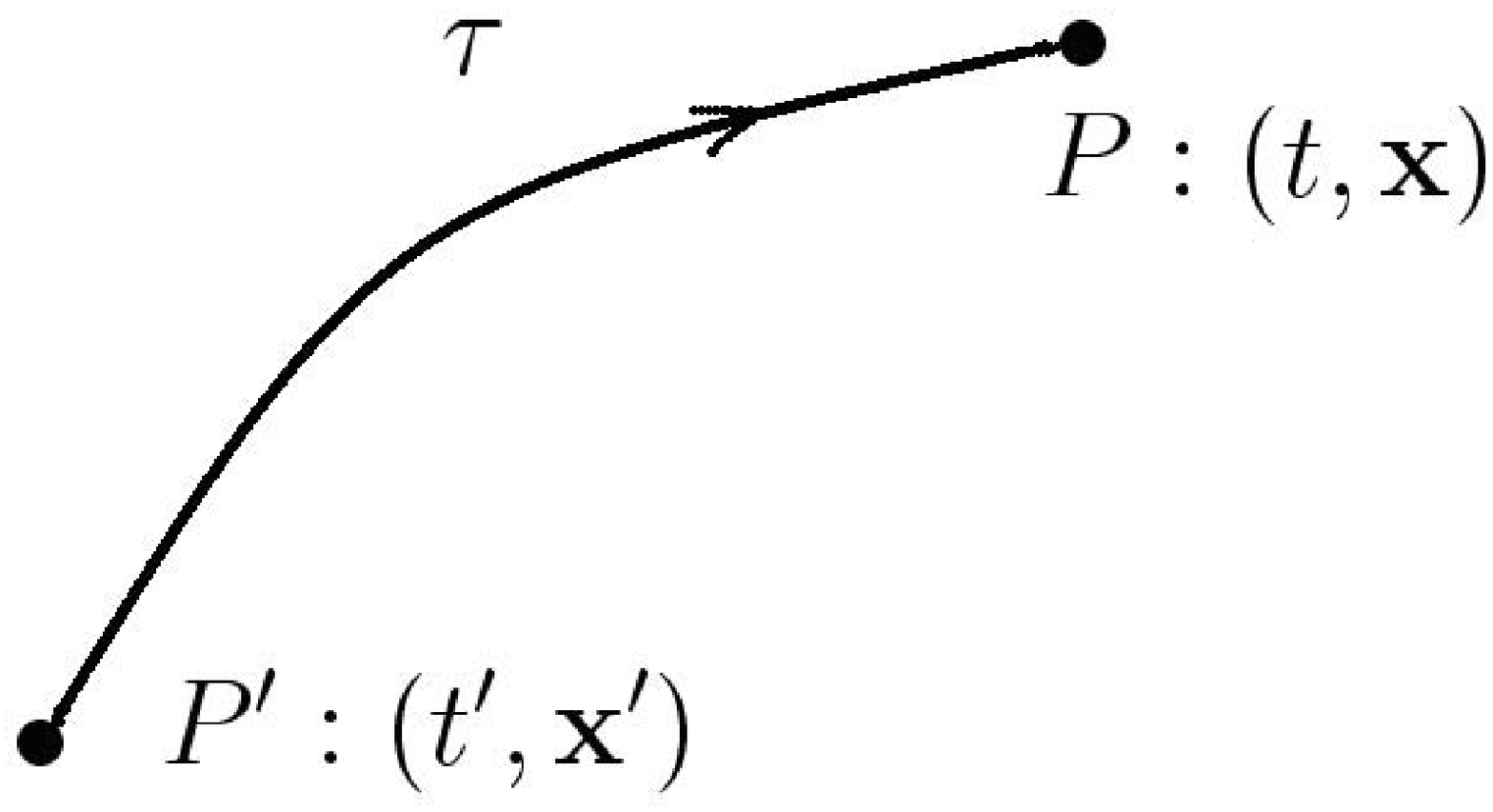} &
\includegraphics[scale=0.35]{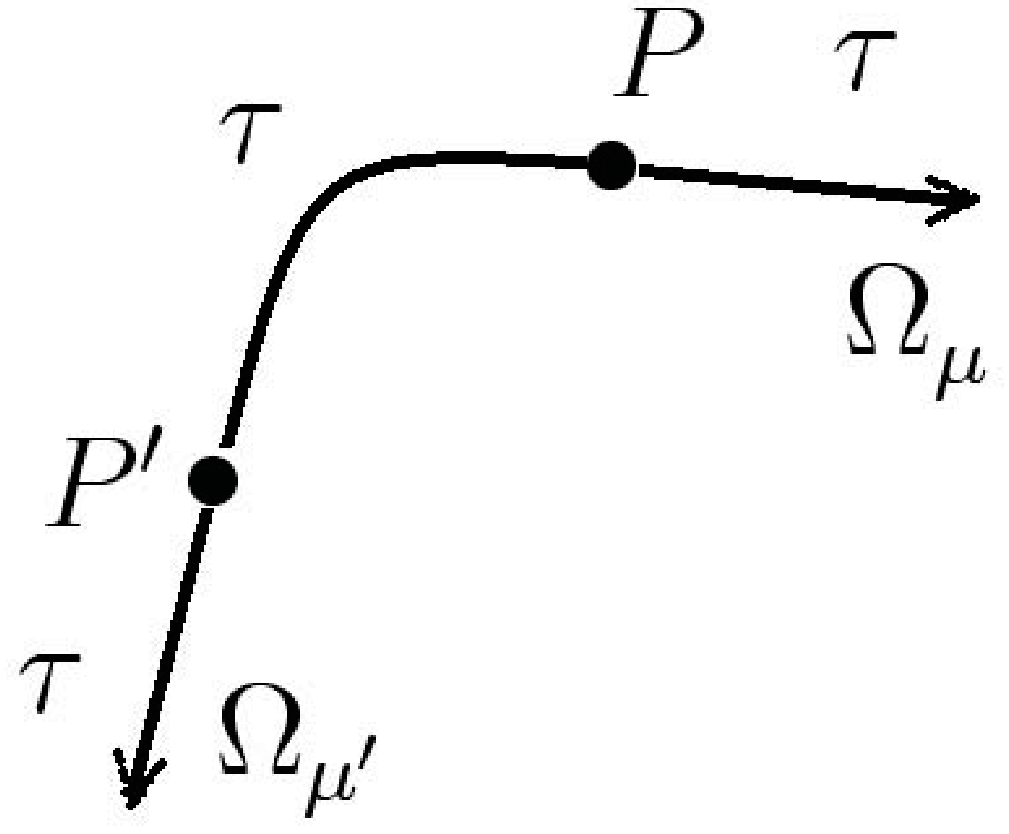}\cr
(a) & (b)\cr
\end{array}
\]
\end{center}
\caption{(a) A timelike geodesic segment with proper time $\tau$ that starts at point $P'$ and ends at point $P$. (b) The derivatives of the world function $\Omega_{\mu'}$ and $\Omega_\mu$ are tangents to the path at $P'$ and $P$, respectively, and have the same length $\tau$ as the geodesic segment.}
\end{figure}

\section{World Function in Certain Conformally Flat Spacetimes}

Let us now consider the world function in our conformally flat spacetimes. To simplify matters, we will 
assume that  $U=U(t)$, where $t \ge 0$ in our convention. As is well known, the geodesics can be obtained from 
\begin{equation}\label{4.1}
 {\cal L}= -\frac{1}{2} \left(\frac{ds}{d\tau}\right)^2\,,
 \end{equation}
where $\tau$ is initially just an affine parameter, but will turn out to be the proper time along a timelike geodesic, once we set ${\cal L}=1/2$ along the geodesic,
\begin{equation}\label{4.2}
\frac{d^2 x^{\mu}}{ds^2}+{^0}\Gamma^\mu_{\alpha \beta}~ \frac{dx^\alpha}{ds} \frac{dx^\beta}{ds} = 0\,.
\end{equation}
Thus
\begin{equation}\label{4.3}
 {\cal L}= \frac{1}{2}\,e^{2U} \,(\dot{t}^2-\delta_{ij}\dot{x}^i\,\dot{x}^j)\,, 
 \end{equation}
where an overdot indicates differentiation with respect to $\tau$.  Hence, 
\begin{equation}\label{4.4}
p_0=\frac{\partial {\cal L}}{\partial \dot{t}}=e^{2U}\dot{t}\,, \qquad   p_i=\frac{\partial {\cal L}}{\partial \dot{x}^i}=-e^{2U}\,\delta_{ij}\,\dot{x}^j\, 
 \end{equation}
and the Euler-Lagrange equations are
\begin{equation}\label{4.5}
\frac{d}{d\tau}\,(e^{2U}\dot{t})=U_{,t}\,e^{2U}(\dot{t}^2-\delta_{ij}\dot{x}^i\,\dot{x}^j)\,, 
 \end{equation}
 \begin{equation}\label{4.6}
 \frac{d}{d\tau}\,(e^{2U}\,\delta_{ij}\,\dot{x}^j)=0\,.
\end{equation}
Equations~\eqref{4.5} and~\eqref{4.6} can be written as
\begin{equation}\label{4.7}
\ddot{t}+U_{,t}\,(\dot{t}^2+\delta_{ij}\dot{x}^i\,\dot{x}^j)=0\,, 
 \end{equation}
 \begin{equation}\label{4.8}
 \delta_{ij}\,\ddot{x}^j+2U_{,t}\,\dot{t}\, \delta_{ij}\,\dot{x}^j =0\,.
\end{equation}
Introducing the Euclidean spatial interval $d\ell$, $d\ell^2=\delta_{ij}\,dx^i\,dx^j$, we have from the integration of these equations for a \emph{timelike geodesic} that
\begin{equation}\label{4.9}
e^{2U}\,\left[\dot{t}^2-\left(\frac{d\ell}{d\tau}\right)^2\right]=1\,, 
 \end{equation}
 \begin{equation}\label{4.10}
\frac{d\ell}{d\tau}=\eta\,e^{-2U}\,, 
\end{equation}
where $\eta$ is a constant of integration. 

Assuming that the timelike geodesic starts at $P'$ and moves forward to $P$ as in Figure 1, we have $\eta \ge 0$ and
\begin{equation}\label{4.11}
\tau=\int_{t'}^{t} \frac{e^{2U(\theta)}}{\sqrt{\eta^2+e^{2U(\theta)}}}\, d\theta\,,
\end{equation}
\begin{equation}\label{4.12}
|\mathbf{x}-\mathbf{x'}|=\eta\,\int_{t'}^{t} \frac{1}{\sqrt{\eta^2+e^{2U(\theta)}}}\, d\theta\,.
\end{equation}
Moreover, it follows from these relations that 
\begin{equation}\label{4.13}
\tau + \eta\, |\mathbf{x}-\mathbf{x'}|=\int_{t'}^{t} \sqrt{\eta^2+e^{2U(\theta)}}\, d\theta\,.
\end{equation}

The parameter $\eta: 0 \to \infty$ must be eliminated between Eqs.~\eqref{4.11} and~\eqref{4.12} to give us the world function $\Omega(x, x') = -\tau^2/2$. We note that $\eta=0$ when there is no movement in space at all, while $\eta=\infty$ corresponds to lightlike motion. Null geodesics are invariant under a conformal transformation; therefore, we see that for $\eta=\infty$, $|\mathbf{x}-\mathbf{x'}|=t-t'$ and $\tau=0$, as expected. 

In many interesting situations examined below, it turns out that $e^{U(\theta)}=C\,\theta^{\,\nu}$ for $\nu>0$ and constant $C>0$.
In this case, Eqs. \eqref{4.11} and \eqref{4.12} become
\begin{equation}\label{4.13a}
\tau=\frac{\eta}{\omega_\nu  }\left[H_\nu\left(\omega_\nu t  \right) -H_\nu\left(\omega_\nu   t'   \right) \right]
\end{equation}
and
\begin{equation}\label{4.13b}
|\mathbf{x}-\mathbf{x'}|=\frac{1}{\omega_\nu  }\left[S_\nu\left(\omega_\nu t  \right) -S_\nu\left(\omega_\nu   t'   \right) \right]\,,
\end{equation}
where
\begin{equation}\label{4.13c}
\omega_\nu \equiv \omega_\nu(\eta) :=\left(\frac{C}{\eta}\right)^{1/\nu}\,,
\end{equation}
and
\begin{equation}\label{4.13d}
H_\nu(x)=\int_0^x \frac{\psi^{2\nu}}{\sqrt{1+\psi^{2\nu}}}d \psi, \qquad S_\nu(x)=\int_0^x \frac{1}{\sqrt{1+\psi^{2\nu}}}d \psi\,.
\end{equation}
Formally, we have for $\nu \ne -1$,
\begin{equation}\label{4.13e}
H_\nu(x)= \frac{1}{\nu+1} \left[  x\sqrt{1+x^{2\nu}}- S_\nu(x)\right]\,.
\end{equation}
From the integral representation of the hypergeometric function given by~\cite{AS}
\begin{equation}\label{4.13f}
F(a, b; c; \zeta)= \frac{\Gamma(c)}{\Gamma(b)\,\Gamma(c-b)}\int_0^1t^{b-1}\,(1-t)^{c-b-1}\,(1-t\,\zeta)^{-a}\,dt\,,
\end{equation}
valid for $Re(c)>Re(b)>0$, with
\begin{equation}\label{4.13g}
t^{1/(2\nu)}\,x=\psi\,,
\end{equation}
we find
\begin{equation}\label{4.13h}
S_\nu(x)=x \, F\left(\frac12 , \frac{1}{2 \nu}; 1+ \frac{1}{2 \nu}; -x^{2\nu}\right)\,.
\end{equation}
It is not in general possible to eliminate $\eta$ between Eqs.~\eqref{4.11} and~\eqref{4.12} explicitly, or in their new forms~\eqref{4.13a} and~\eqref{4.13b}, to obtain an \emph{explicit}  expression for the world function. To illustrate this situation, we consider some special cases. 

\subsection{Minkowski Spacetime}

Let $U=0$, as in Minkowski spacetime. Then, 
\begin{equation}\label{4.14}
\tau = \frac{t-t'}{\sqrt{\eta^2+1}}\,, \qquad  |\mathbf{x}-\mathbf{x'}|= \frac{\eta\,(t-t')}{\sqrt{\eta^2+1}}\,.
\end{equation}
Here $\eta=v/\sqrt{1-v^2}$, where $v$ is the speed of uniform motion along the straight line from $P'$ to $P$. Thus, eliminating $\eta$, we find Eq.~\eqref{OmegaMink1} for $^M\Omega$ of Minkowski spacetime. 

\subsection{de Sitter Spacetime}

Let $e^U=1/(\lambda \, t)$, where $\lambda:=\sqrt{\Lambda/3}$ and $t\ge0$, as in de Sitter spacetime~\cite{Ru}. Then, Eqs.~\eqref{4.11} and~\eqref{4.12} imply that for $u:=\eta\,\lambda\,t$ and $u':=\eta\,\lambda\,t'$,
\begin{equation}\label{4.15}
e^{\lambda\, \tau}= \frac{u}{u'}\,\frac{1+\sqrt{1+u'^2}}{1+\sqrt{1+u^2}}\,, 
\end{equation}
\begin{equation}\label{4.16}
\eta\,\lambda\, |\mathbf{x}-\mathbf{x'}|=\sqrt{1+u^2}-\sqrt{1+u'^2}\,. 
\end{equation}
Eliminating $\eta$ in this case leads to $\Omega(x, x') = -\tau^2/2$, where 
\begin{equation}\label{4.17}
Q=\cosh{(\lambda\,\tau)}\,, \qquad Q=\frac{t^2+t'^2-|\mathbf{x}-\mathbf{x'}|^2}{2t\,t'}\,.\
\end{equation}
Here, $Q$ satisfies
\begin{equation}\label{4.18}
g^{\mu \nu}\,Q_{,\mu}\,Q_{,\nu}=\lambda^2 (1-Q^2)\,.
\end{equation}
More specifically, we have $Q\ge 1$,  
\begin{equation}\label{4.18a}
\lambda\,\tau=\ln{(Q+\sqrt{Q^2-1})}\,
\end{equation}
and
\begin{equation}\label{4.18b}
\Omega=-\frac{1}{2\,\lambda^2}\,\ln^2{(Q+\sqrt{Q^2-1})}\,.
\end{equation}

Using relations~\eqref{4.17} and~\eqref{4.18a}, it is straightforward to check that over an infinitesimal interval with $t-t'=dt >0$ and $|\mathbf{x}-\mathbf{x'}|=|d\mathbf{x}|$, we have, as expected,  
\begin{equation}\label{4.18c}
d\tau= \frac{1}{\lambda\,t}\,\sqrt{dt^2- d\mathbf{x}^2}\,
\end{equation}
for de Sitter spacetime. 

\subsection{FLRW Spacetime with Stiff Equation of State} 

Next, we consider the case of a flat FLRW universe with a stiff equation of state $p=\rho$, as discussed in Ref.~\cite{Ro}. Here, $e^U=\sqrt{\beta\,t}$ and $\rho=p=3/(32 \pi G \beta t^3)$. From Eq.~\eqref{4.12}, we find
\begin{equation}\label{4.19}
\beta\,|\mathbf{x}-\mathbf{x'}|= 2 \eta\, (W-W')\,, \qquad W:=\sqrt{\eta^2+\beta\,t}\,.
\end{equation}
Squaring this relation twice and defining 
\begin{equation}\label{4.20}
\tau_M^2:=(t-t')^2-|\mathbf{x}-\mathbf{x'}|^2\,,
\end{equation}
we find, after some algebra, that 
\begin{equation}\label{4.21}
\eta^2=\frac{\beta}{4\,\tau_M^2}\,|\mathbf{x}-\mathbf{x'}|^2\,\left(t+t'+\sqrt{4\,t\,t'+|\mathbf{x}-\mathbf{x'}|^2}\right)\,,
\end{equation}
which properly diverges for null motion $(\tau_M=0)$. Next, Eq.~\eqref{4.11} implies that 
\begin{equation}\label{4.22}
\frac{3}{2}\,\tau +\eta\,|\mathbf{x}-\mathbf{x'}|=t\,W-t'\,W'\,.
\end{equation}
Substituting for $\eta$ in this equation, we eventually find Roberts's explicit expression for the world function in this case~\cite{Ro}. 

\subsection{Einstein-de Sitter Universe}

Finally, we consider the case of a flat FLRW universe with $p=0$ and  $\rho=3/(2\pi G b^4 t^6)$, where $b=1/(3t_0)$ and $t_0$ is the age of the universe in this model. Thus the present energy density of matter  $\rho_0$ is given by $6\pi G \rho_0 t_0^2 = 1$. In this case, $e^{U}=b^2 t^2$ and only an implicit form of the world function is possible. 

\section{The $1+1$ case: An illustrative example}

We have seen above that, even in the simple case of conformally flat spacetimes, the various steps leading to the computation of the main objects of interest for NLG are rather involved and it is difficult to obtain closed form expressions for the world function $\Omega$ and hence for the nonlocal tensor $N_{\alpha\beta\gamma}$.

It is interesting to study then the simplified situation of a conformally flat spacetime in $1+1$ dimensions associated with coordinates $(t,x)$, i.e., with metric
\begin{equation}\label{5.1}
ds^2=e^{2U(t,x)}(-dt^2+dx^2)\,,\qquad \sqrt{-g}=e^{2U(t,x)}.
\end{equation}
Indeed, any 2D spacetime is conformally flat~\cite{Ste}; hence, it can be represented in the form of Eq.~\eqref{5.1}. In this case, the Ricci tensor, scalar curvature and Einstein tensor are given by
\beq \label{5.2}
{}^0R_{\mu\nu}=-\eta_{\mu \nu}\,U^{\alpha}{}_{\alpha}\,,\qquad
{}^0R=-2\,e^{-2U}\,U^{\alpha}{}_{\alpha}\,,\qquad
{}^0G_{\mu\nu}=0\,.
\eeq 
We assume that in 2D spacetime, $\Lambda=0$ and $T_{\mu\nu}=0$; therefore, any 2D spacetime satisfies the GR field equations, namely, ${}^0G_{\mu\nu}=0$. Moreover, the field equations of the NLG field in $1+1$ dimensions reduce to
\beq \label{5.3}
{\mathcal N}_{\mu\nu}=Q_{\mu\nu}
\eeq
where $Q_{\mu\nu}$,  defined to be traceless in 4D, is not in general traceless in 2D.
A straightforward calculation shows that
\begin{equation}\label{5.4}
{\mathcal N}_{\mu\nu}=e^{-U}g_{\nu\alpha}  \partial_\beta (e^{U} N^{\alpha\beta}{}_\mu)\,.
\end{equation}
The connection of $N_{\alpha \beta \gamma}$ with the kernel of NLG follows from Eq.~\eqref{2.10}, which in this case can be written as
\begin{eqnarray}\label{5.4a}
\frac12\, N^{01}{}_0 &=& \int T \Delta\, {\cal K}\, e^{2U(t',x')}dt' dx'\,, \nonumber\\
\frac12\, N^{01}{}_1 &=& \int X \Delta\, {\cal K}\, e^{2U(t',x')}dt' dx'\,. 
\end{eqnarray}
Here, we have defined
\begin{eqnarray} \label{5.4b}
T(x,x')&:=&U_{0'}\Omega_{01'}-U_{1'}\Omega_{00'}\,, \nonumber\\
X(x,x')&:=& U_{0'}\Omega_{11'}-U_{1'}\Omega_{10'}\,, \nonumber\\
\Delta(x,x')&:=& \Omega^{00'}\Omega^{11'}-\Omega^{10'}\Omega^{01'}\,.
\end{eqnarray}
Furthermore,  ${\cal K}$ is  the causal kernel of the theory and hence nonzero only over the past light cone; therefore, ${\cal K}$ is  proportional to the Heaviside unit step function, ${\cal K}\propto \Theta(t-t'-|x-x'|)$. 

To explore the implications of the field equations in this 2D case, we note that
\beq \label{5.5}
C_{\alpha\beta\gamma}=2U_{[\alpha} g_{\beta]\gamma}\,,
\eeq
and
\beq \label{5.6}
Q_{\mu\nu}=g_{\nu\alpha}(U_\mu N^{\alpha\rho}{}_{\rho}-U_\sigma N^{\alpha\sigma}{}_{\mu})-\frac12 g_{\mu\nu}U_\alpha N^{\alpha\rho}{}_{\rho}\,.
\eeq
Hence, we find $Q_{01}=Q_{10}=0$ and
\beq \label{5.7}
Q_{00}=-\frac12 e^{2U}(U_0 N^{01}{}_{1}-U_1 N^{01}{}_{0})=-Q_{11}\,.
\eeq
Moreover, the trace of $Q_{\mu\nu}$ is nonzero and it is given by
\beq \label{5.8}
Q^\alpha{}_\alpha=U_0 N^{01}{}_{1}-U_1 N^{01}{}_{0}\,.
\eeq
The NLG field equations ${\mathcal N}_{\mu\nu}=Q_{\mu\nu}$ 
imply ${\mathcal N}_{01}={\mathcal N}_{10}=0$; that is,
\beq \label{5.9}
\partial_0 (e^U N^{01}{}_0)=\partial_1 (e^U N^{01}{}_1)=0\,.
\eeq
Moreover, $\mathcal{N}_{00}=Q_{00}$ and $\mathcal{N}_{11}=Q_{11}$ imply, respectively,
\begin{eqnarray} \label{5.10}
\partial_1 (e^U N^{01}{}_0) &=& \frac12 e^U(U_0 N^{01}{}_{1}-U_1 N^{01}{}_{0})\,\nonumber\\
\partial_0 (e^U N^{01}{}_1) &=&-\frac12 e^U(U_0 N^{01}{}_{1}-U_1 N^{01}{}_{0})\,.
\end{eqnarray}

It follows from Eqs.~\eqref{5.9}  that 
\beq \label{5.11}
\frac12\, e^U N^{01}{}_0=f(x)\,,\qquad \frac12\,e^U N^{01}{}_1=h(t)\,,
\eeq
where $f(x)$ and $h(t)$ are two arbitrary functions.
Furthermore, from Eqs.~\eqref{5.10} we find
\beq \label{5.12}
\partial_1 (e^U N^{01}{}_0)+\partial_0 (e^U N^{01}{}_1)=0\,;
\eeq
that is,
\beq \label{5.13}
\partial_x f(x)=-\partial_t h(t)= \alpha\,,
\eeq
where $\alpha$ is a constant and we have in general 
\beq \label{5.14}
f(x)=\alpha x + f_0\,,\qquad h(t)=-\alpha t+ h_0\,,
\eeq
where $f_0$ and $h_0$ are integration constants. At this point, we encounter two distinct possibilities, which will now be discussed in turn.

\subsection{$\alpha=0$}

In this case, Eqs.~\eqref{5.10} imply that 
\beq \label{5.14a}
h_0\, \partial_t U - f_0\, \partial_xU = 0\,,
\eeq
which means that $U$ is simply a function of $\varphi:=f_0\, t + h_0\, x$ and the 2D spacetime thus possesses a Killing vector field $h_0\, \partial_t  - f_0\, \partial_x$. This Killing vector is timelike, specelike or null depending upon whether $f_0^2-h_0^2$ is negative, positive or zero, respectively. 

The kernel of NLG in this case must be determined from Eqs.~\eqref{5.4a}, namely, 
\begin{eqnarray}\label{5.14b}
 f_0\,e^{-U(\varphi)} &=& \int T \Delta\, {\cal K}\, e^{2U(\varphi')}dt' dx'\,, \nonumber\\
 h_0\,e^{-U(\varphi)} &=& \int X \Delta\, {\cal K}\, e^{2U(\varphi')}dt' dx'\,. 
\end{eqnarray}

\subsection{$\alpha \ne 0$}

In this case, $f_0$ and $h_0$ essentially amount to  a constant spacetime  translation of coordinates  $x$ and $t$, respectively, and therefore can be set equal to zero with no loss in generality. It then follows from
 Eqs.~\eqref{5.10} that
\beq \label{5.15}
x^\mu\, \partial_\mu U =-2\,.
\eeq
We can use Euler's theorem on homogeneous functions to conclude that
\beq \label{5.16}
e^U=\frac{1}{t^2}\,\phi\left( \frac{x}{t}\right)\,,
\eeq
where $\phi$ is an arbitrary smooth function of the similarity variable $x/t$.
That is, $e^U$ should be a homogeneous function of $t$ and $x$ of degree $-2$. The 2D spacetime of NLG in this case admits a homothetic Killing vector, $K=t\,\partial_t+x\,\partial_x$, such that
\beq \label{5.16a}
^0\nabla_{\alpha} K_{\beta}\, +\, ^{0}\nabla_{\beta} K_{\alpha}=- 2\,g_{\alpha\beta}\,.
\eeq

As before, the kernel of NLG in this case must be determined from Eqs.~\eqref{5.4a}, namely, 
\begin{eqnarray}\label{5.17}
 \alpha\, e^{-U} x &=& \int T \Delta\, {\cal K}\, e^{2U(t',x')}dt' dx'\,, \nonumber\\
- \alpha\, e^{-U} t &=& \int X \Delta\, {\cal K}\, e^{2U(t',x')}dt' dx'\,. 
\end{eqnarray}
These equations are compatible provided
\beq \label{5.18}
 \int (t\,T +x\,X)\Delta\, {\cal K}\, e^{2U(t',x')}dt' dx'=0\,.
\eeq

\subsection{2D de Sitter Spacetime}

Finally, it is interesting to consider in some detail the simple case of 2D de Sitter spacetime that satisfies the field equations of NLG. This case corresponds to $\alpha=0$, $f_0=\lambda$, $h_0=0$ and $U=-\ln{(\lambda\,t)}$. It remains to determine a kernel $\mathcal{K}(x, x')$ that satisfies relations~\eqref{5.14b} in this case. To see what this entails, we note that these relations reduce to 
\begin{eqnarray}\label{5.19}
\int \frac{1}{t'^3}\, (\Omega_{01'} \, \Delta) \,\mathcal{K}(x, x') \,dt'dx' &=& -\lambda^4 \,t\,, \nonumber\\
\int \frac{1}{t'^3}\, (\Omega_{11'} \, \Delta) \,\mathcal{K}(x, x') \,dt'dx' &=& 0\,.
\end{eqnarray}

The world function can now be employed to  determine $\Omega_{\alpha \alpha'}$ and $\Delta$. We recall from Eqs.~\eqref{4.17} and~\eqref{4.18b} that 
\beq\label{5.20}
\Omega= -\frac{1}{2\lambda^2}\ln^2 (Q+\sqrt{Q^2-1})\,, \qquad    Q=\frac{t^2+t'^2-(x-x')^2}{2tt'}\,.
\eeq
The mixed derivatives of the world function can be conveniently written as 
\begin{equation}\label{5.21}
\Omega_{\alpha \alpha'}=\Upsilon\,\left( \Phi\, A_{\alpha \alpha'}+ B_{\alpha \alpha'} \right)
\end{equation}
where
\beq \label{5.22}
\Upsilon^{-1} := \lambda^2 \, t^2 t'^2\, (Q^2-1)\,, \qquad \Phi :=\frac{\ln{(Q+\sqrt{Q^2-1})}}{\sqrt{Q^2-1}}\,.
\eeq
Moreover, 
\beq \label{5.23}
 A_{\alpha \alpha'}:= -Q\, B_{\alpha \alpha'} - t^2 t'^2\, (Q^2-1)\,Q_{\alpha \alpha'}\,
 \eeq
and 
\beq \label{5.24}
 B_{\alpha \alpha'}:=  - t^2 t'^2\, Q_{\alpha}\,Q_{\alpha'}\,,
 \eeq
where $Q_{\alpha} :=\partial_{\alpha}Q$ and $Q_{\alpha \alpha'} :=\partial_{\alpha}\partial_{\alpha'}Q$.
It follows from a detailed calculation  that 
\begin{eqnarray}\label{5.25}
A_{t t'}=-B_{x x'}=-(x-x')^2\,, \nonumber \\
A_{t x'} = - B_{t' x}=(x-x')(Q\, t-t')\,, \nonumber  \\ 
A_{x t'} = -B_{x' t} =-  (x-x')(Q\,t'-t)\,, \nonumber \\ 
A_{x x'} = -B_{t t'} = (Q\,t-t')(Q\,t'-t)\,.
\end{eqnarray}
These results reveal a simple expression for $\Delta$, namely, 
\beq \label{5.26}
\Delta= -\lambda^4 t^2t'^2\, \Phi\,
\eeq
in  2D de Sitter spacetime. It is clear that even in this simple case, the determination of a causal kernel  $\mathcal{K}(x, x')$ in conformity with relations~\eqref{5.19} is a daunting task that is beyond the scope of this work.

\section{Discussion}

Nonlocal gravity is a classical generalization of Einstein's general relativity in which nonlocality is due to the gravitational memory of past events. The gravitational field is local, but satisfies integro-differential field equations. History dependence is introduced into the theory through a scalar causal constitutive kernel. It is not known how this kernel should be determined from basic principles;  perhaps a more comprehensive future theory is needed  to fix the kernel in 4D. In the absence of such a theory,  we can in principle use observational data regarding dark matter to determine the kernel.  In this paper, we have studied the field equations of nonlocal gravity within the framework of general relativity and explored some of their consequences in the simple case  of  conformally flat spacetimes. In these spacetimes, in particular, we have investigated the world function. To render the analysis more tractable, we have considered the implications of nonlocal gravity in 2D spacetimes.  In 4D spacetime, the nonlocal kernel $\mathcal{K}$ must ultimately be determined from the observational data~\cite{NL5, NL8, NL9}; however, in a 2D spacetime, $\mathcal{K}$ satisfies certain integral equations. 
While any 2D spacetime is conformally flat and satisfies Einstein's source-free field equations, nonlocal gravity imposes the additional requirement that the spacetime should contain either an isometry or a homothety.  We work out in detail  the case of 2D de Sitter spacetime and show that even in this simple manifold the determination of the nonlocal kernel is nontrivial and remains a task for the future.   

\section*{Acknowledgments}
D.B. thanks ICRANet for partial support.

\end{document}